
\magnification=\magstep1
\looseness=2
\tolerance 1000
\def\ref{\par\noindent\hangindent 12pt}

\def\msn{\par\nobreak\noindent}
\def\bsn{\bigbreak\goodbreak\noindent}
\def\noi{\noindent}
\def\etal{{\it et al.}\ }
\def\refset{\parindent=0pt\hangafter=1\hangindent=1em}

\hsize=6.25 truein
\vsize=9.0 truein
\baselineskip=25pt plus1pt minus1pt
\parskip=2pt plus1pt minus1pt
\vfill
\centerline{THE MOTIONS OF CLUSTERS AND GROUPS OF GALAXIES}
\vskip 2cm
\centerline{Neta A. Bahcall,$^1$ Mirt Gramann,$^{1,2}$ and
Renyue Cen$^1$}
\bsn
\centerline{$^1$ Princeton University Observatory, Princeton, NJ 08544}
\centerline{$^2$ Tartu Astrophysical Observatory, T\~oravere, Estonia}
\vskip 2cm
\centerline {\hskip 2cm Received:\hrulefill; Accepted:\hrulefill \hskip 2cm}
\vskip 6cm
\centerline{Submitted to the {\it Astrophysical Journal}}
\vskip 1cm
\centerline{February 2, 1993}
\par\vfill\eject

\centerline {ABSTRACT}
\msn
The distributions of peculiar velocities of rich clusters and of groups
of galaxies are investigated for different cosmological models and are
compared with observations. Four cosmological models are studied:
standard ($\Omega=1$) CDM, low-density CDM, HDM ($\Omega=1$), and
PBI. We find that rich clusters of galaxies exhibit a Maxwellian
distribution of peculiar velocities in all models, as expected from a
Gaussian initial density fluctuation field. The clusters appear to
be a fundamental and efficient tracer of the large-scale
velocity-field. The cluster 3-D velocity distribution is generally
similar in the models: it peaks at $v \sim 500$ km s$^{-1}$, and
extends to high cluster velocities of $v \sim 1500$ km s$^{-1}$.
Approximately 10\% of all model rich clusters move with high peculiar
velocities of $v \ge 10^3$ km s$^{-1}$. The highest velocity
clusters frequently originate in dense superclusters. The model
velocity distributions of rich clusters are compared with the model
velocity distributions of small groups of galaxies, and of the total
matter. The group velocity distribution is, in general, similar
to the velocity distribution of the rich clusters. In all but the
low-density CDM model, the mass exhibits a longer tail of high
velocities than do the clusters. This high-velocity tail originates
mostly from the high velocities that exist {\it within} rich clusters.
The low-density CDM model exhibits Maxwellian velocity distributions
for clusters, for groups, and for matter that are all similar to each
other.

The model velocity distributions of groups and clusters of galaxies
are compared with observations. The data are generally consistent with
the models, but exhibit a somewhat larger high-velocity tail, to
$v_r \sim 3000$ km s$^{-1}$. While this high-velocity tail is similar
to the HDM model predictions, the data are consistent with the other
models studied, including the low-density CDM model, which best fits
most other large-scale structure observations. The observed velocity
distribution is consistent with a Gaussian initial density
fluctuation field.

{\it Subject headings: cosmology: theory -- galaxies: clustering}
\par\vfill\eject
\centerline{1. INTRODUCTION}
\msn
Clusters of galaxies are an efficient tracer of the large-scale
structure of the universe. Measurements of the strong correlation
function of cluster of galaxies (Bahcall \& Soneira 1983; Klypin
\& Kopylov 1983; Postman \etal 1992; Peacock \& West 1992;
Bahcall \& West 1992; Dalton \etal 1992; Nichol \etal 1992),
and the superclustering properties of clusters (Bahcall \& Soneira
1984, Postman \etal 1992, Einasto \etal 1994, Rhoads \etal 1994)
provided some of the first evidence for the existence of organized
structure on large scales. The existence of large-scale structure
produces gravitationally-induced large-scale peculiar velocities
of galaxies, of groups of galaxies, and of clusters of galaxies.
Large-scale peculiar motions of galaxies have been detected
(Rubin \etal 1976, Dressler \etal 1987, Burstein \etal 1987,
Faber \etal 1989); combined with observations of the large-scale
distribution of galaxies, these peculiar velocities provide
important constraints on cosmological models
(Bertschinger \& Dekel 1989; Kofman \etal 1993;
Croft \& Efstathiou 1993; Dekel 1994 and references therein).
The motions of clusters of galaxies have also been investigated.
Bahcall \etal (1986) suggested the possible existence of large
peculiar velocities of clusters in some dense superclusters,
with $v_r \sim 10^3$ km s$^{-1}$. Aaronson \etal (1983), Faber \etal
(1989), and Mould \etal (1991, 1993) find, using direct distance
indicators, clusters with similarly high peculiar velocities,
$v_r \sim 10^3$ km s$^{-1}$, in some regions. Lauer and Postman
(1993) have recently reported a large bulk-flow of relatively
nearby rich clusters of galaxies, with $v_r \sim 800$ km s$^{-1}$.

What is the role of rich clusters of galaxies in mapping the velocity
field of the universe? Do rich clusters provide a useful tracer of
the large-scale velocity field? In this paper, we study the velocity
distribution of rich clusters of galaxies in several popular
cosmological models. We also compare the velocity distribution of
rich clusters with that of groups and of the total matter for each
of the models. We conclude that rich clusters of galaxies provide
efficient tracers of large-scale velocities. We compare the model
expectations with observations of group and cluster velocities.
We find that the observational data on group and cluster velocities
are consistent with most of the predictions of these popular
cosmological models.

We describe the model simulations in \S 2. In \S 3 we present the
distribution of peculiar velocities of rich clusters of galaxies in
the models. We compare in \S 4 the rich-cluster velocity distribution
with that of small groups of galaxies and of the total matter
distribution. We present in \S 5 a fit of the cluster velocity
distribution to a Maxwellian. In \S 6 we investigate the velocity
distribution of clusters within dense superclusters in the
models. We compare the cluster and group velocity distribution with
observations in \S 7, and summarize the main conclusions in \S 8.

\bsn
\centerline{2. MODEL SIMULATIONS}
\msn

A large-scale Particle-Mesh code with a box size of $400$h$^{-1}$ Mpc
is used to simulate the evolution of the dark matter in different
cosmological models. The simulation box contains $500^3$ cells and
$250^3 = 10^{7.2}$ dark matter particles. The spatial resolution
is $0.8$h$^{-1}$ Mpc. Details of the simulations are discussed
in Cen (1992).

Four cosmological model simulations are performed: Standard Cold
Dark-Matter (CDM) model ($\Omega=1$); low-density CDM model
($\Omega=0.3$); Hot Dark-Matter model (HDM, $\Omega=1$); and
Primeval Baryonic Isocurvature model (PBI, $\Omega=0.3$). The
specific model parameters are summarized below. These parameters
include the matter density, $\Omega$; the cosmological constant
contribution, $\Omega_{\Lambda}$; the Hubble constant
(in units of $H_O=100$h km s$^{-1}$ Mpc$^{-1}$); and the
normalization of the mass fluctuations on $8$h$^{-1}$ Mpc scale,
$\sigma_8$. The models are normalized to the large-scale microwave
background anisotropy measured by COBE (Smoot \etal 1992).
(The HDM model is $2\sigma$ above the COBE normalization, in
order to minimize late galaxy formation; Cen and Ostriker 1992.)
The parameters of the PBI and low-density CDM models are described
in details by Cen, Ostriker, \& Peebles (1993) and
Cen, Gnedin, \& Ostriker (1993). Both the CDM and HDM models use
the adiabatic density perturbation power spectra given by
Bardeen \etal (1986). The models and their parameters are summarized
in Table 1.

Clusters are selected in each simulation using an adaptive linkage
algorithm following the procedure described by Suto, Cen \& Ostriker
(1992) and Bahcall \& Cen (1992).  The cluster mass threshold, within
$r=1.5$h$^{-1}$ Mpc of the cluster center, is selected to correspond
to a number density of clusters comparable to the observed density
of rich ($R \ge 1$) clusters, $n_{cl} \sim
6 \times 10^{-6}$ h$^3$ Mpc$^{-3}$
(Bahcall \& Cen 1993). When analyzing smaller groups of galaxies
(\S 4), the threshold is appropriately reduced to correspond to
the observed number density of small groups of galaxies,
$n_{gr} \sim 2-5 \times 10^{-4}$ h$^{3}$ Mpc$^{-3}$
(Ramella \etal 1989; Bahcall \& Cen 1993).

A total of $\sim 400$ rich clusters of galaxies and $\sim 3 \times 10^4$
groups are obtained in each of the $400$h$^{-1}$ Mpc simulation models.
The three-dimensional peculiar velocity of each of these clusters
(or groups), relative to the co-moving cosmic background frame, is
determined from the simulation; these velocities are used in the
analyses described below. The analysis is carried out independently
for rich clusters (\S 3), groups (\S 4), and matter (\S 4), in order
to compare the velocity distributions of the different systems.

\bsn
\centerline {3. PECULIAR VELOCITIES OF CLUSTERS OF GALAXIES}
\msn

How fast do rich clusters of galaxies move? We present in Figure~1
the results for the integrated peculiar velocity distribution of
rich clusters for each of the models studied. The integrated
velocity distribution represents the probability distribution, or
the normalized number density, of clusters with peculiar velocities
larger than $v$, $P(>v)$. The velocity $v$ refers to the
three-dimensional peculiar velocity of the cluster relative to
the cosmic background frame.

Two results are immediately apparent from Figure~1. First, the
cluster velocity distribution, which exhibits a moderate fall-off
at high velocities ($v \sim 500-1500$ km s$^{-1}$), and a
leveling-off at small velocities ($v \le 500$ km s$^{-1}$),
is similar in all models (with PBI exhibiting somewhat lower
velocities than the other models). Second, a significant fraction of
all model rich clusters ($\sim 10$\%) exhibit high peculiar
velocities of $v \ge 10^3$ km s$^{-1}$ (somewhat lower for PBI).
The tail of the cluster velocity distribution reaches
$\sim 1500 - 2000$ km s$^{-1}$.

The differential velocity distribution of the clusters, $P(v)$
(i.e., the normalized number density of clusters with peculiar
velocity in the range $v \pm dv$, per unit $dv$, as a function
of $v$), is presented in Figure~2. The cluster velocity
distribution is similar for all but the PBI model, as is
expected from Figure~1. Specifically, the high and low density CDM
models yield very similar results. The HDM results are similar,
with a somewhat stronger tail at high velocities.
(The tail decreases for a lower $\sigma_8$ normalization; Table~1.)
The PBI model exhibits a shift to lower velocities than the other
models. The specific shape of the velocity distribution is also of
interest. The cluster velocities peak at $v \sim 500$ km s$^{-1}$
in all models ($\sim 300$ km s$^{-1}$ for PBI). The distribution
decreases rapidly at lower velocities, and decreases more slowly
at higher velocities. Approximately 30\% of all clusters exhibit
peculiar velocities in the range $v\approx 500 \pm 100$ km s$^{-1}$,
independent of the model. Approximately 10 - 20\% of all clusters
have peculiar velocities in the range $v \approx 1000 \pm
200$ km s$^{-1}$ (lower for PBI).

The existence of a high velocity tail with velocities of
$\sim 1500 - 2000$ km s$^{-1}$ is apparent in Figures 1 and 2.

What causes the cluster velocity distribution to peak at
$500$ km s$^{-1}$ and exhibit the shape seen above?  What is the
origin of the large peculiar velocities of clusters,
$v \ge 10^3$ km s$^{-1}$?  We address these questions in
\S 5 and 6.

\bsn
\centerline {4. PECULIAR VELOCITIES OF CLUSTERS, GROUPS, AND MATTER}
\msn

In this section we compare the velocity distribution of rich clusters
of galaxies, discussed above, with the velocity distributions of poorer
systems, such as poor clusters and small groups of galaxies, as well
as with the velocity of the underlying matter distribution.

Groups of galaxies are selected as described in \S 2, using an
adaptive linkage algorithm, and a mass threshold corresponding to
a total number density of $n_{gr} \sim 4 \times 10^{-4}$ groups
Mpc$^{-3}$; this density is comparable to the observed density of
low-threshold small groups of galaxies (Ramella \etal 1989;
Bahcall \& Cen 1993). About $\sim 3 \times 10^4$ such groups
are identified in each of the simulations. The velocity
distribution function for groups is then determined as
described in \S 3.

The velocity distribution of all the matter (i.e., the velocities
of all the dark-matter particles) in each simulation is also determined.
The matter velocity distribution may represent the closest
approximation to the velocity distribution of the galaxies.

The differential and integrated velocity distributions of the
groups and of all the matter in each of the four models are
presented in Figures 3(a-d) and 4(a-d). In each sub-figure
(a to d), the velocity distribution of the rich clusters, groups,
and matter are compared with each other for a given model.

The differential velocity distribution of the clusters, groups,
and matter (Figure 3) reveals - for all models - a peaked distribution.
The distributions typically peak at peculiar velocities around
$500$ km s$^{-1}$, and exhibit an extended tail to high velocities
of $\ge 10^3$ km s$^{-1}$.  A comparison of the velocity distributions
in the different models yields interesting results. First, in
the low density models (especially CDM $\Omega=0.3$, and, to
a lesser extent, PBI), the velocity distributions of the
clusters, groups, and matter are all similar to each other.
This is clearly seen in the differential (Figures 3b, 3d) and
in the integrated (4b, 4d) velocity distributions. In the PBI model,
the matter distribution is shifted slightly to higher velocities than
the groups and clusters. The $\Omega=1$ models (CDM, HDM) reveal
a significant difference between the rich clusters and the
matter velocity distributions; the matter exhibits a stronger
tail to large velocities than do the rich clusters (Figures
3a, 3c; 4a, 4c). The effect is seen most clearly in the
integrated function, Figure~4. The peak of the matter velocity
distribution ($\Omega=1$ models) is also shifted to a somewhat
higher velocity ($\sim 700$ km s$^{-1}$) than for the rich
clusters ($\sim 500$ km s$^{-1}$). The velocity distributions
of small groups are typically intermediate between the velocity
distributions of the clusters and of the matter.

What is the origin of the high-velocity tail in the velocity
distribution of matter in the $\Omega=1$ models? The matter
(or galaxy) velocity distribution includes the velocities of
the matter particles (or galaxies) {\it within} rich clusters.
The 3-D velocity dispersion within rich clusters is very high for
$\Omega=1$ models such as CDM (with $\sigma_8=1$), reaching
$\sim 3000$ km s$^{-1}$ (e.g., Bahcall \& Cen 1992).
These large velocities provide the main contribution to the
high velocity tails seen in the $\Omega=1$ models (as well
as the excess seen in the matter velocity of the PBI model).
Other than this additional small-scale mass contribution, the
galaxies and clusters appear to trace the same velocity
distribution on large scales.

The low-density CDM model does not show a high velocity tail for the
matter distribution. The model velocity distributions for groups,
for rich clusters, and for all the matter are similar
(Figures 3b, 4b). In this model, the internal velocity-dispersions
in clusters must therefore be similar to (or smaller than) the
high velocity tail of the clusters themselves.  This is indeed so;
the 3-D velocity dispersion in rich clusters in the
$\Omega=0.3$ CDM model, $\sim 1500$ km s$^{-1}$ (consistent with
cluster observations; Bahcall \& Cen 1992), is comparable to the
tail of the cluster peculiar velocities $v \sim 1500$ km s$^{-1}$
(Figures 3b, 4b). This model is consistent with most current
observations of large-scale structure, including the galaxy and
the cluster correlation functions, the power-spectrum of galaxies,
the small-scale peculiar velocities of galaxies, and the observed
mass-function of galaxy clusters (Maddox \etal 1989;
Bahcall \& Cen 1992; Efstathiou \etal 1992; Ostriker 1993;
Kofman \etal 1993). Observational determination of the velocity
distribution of both rich clusters and of galaxies, and their
comparison with the predictions of Figures 3b and 4b, will provide
a further test of this currently promising model (see \S 7).

\bsn
\centerline {5. MAXWELLIAN FITS OF THE VELOCITY DISTRIBUTION}
\msn

The velocity distributions of clusters, of groups, and of matter
were fitted with Maxwellian distributions
$P(v) \propto v^2 \exp(- v^2/2 \sigma^2)$. The velocity dispersion
for each case ($\sigma$) was determined from the model simulation.
The results are presented in Figures 5 and 6. The velocity
distributions of clusters and of all the matter (dark solid
and dotted lines, respectively) are compared with the Maxwellian
distribution (faint lines).  The differential (Figure 5) and the
integrated (Figure 6) velocity distributions are shown for the
$\Omega=1$ and $\Omega=0.3$ CDM models.

The velocity distribution of clusters of galaxies is well
represented by a Maxwellian for {\it all} models (CDM, HDM, and PBI).
(The small deviation suggested at the highest velocities may be due
to clusters that are located in dense superclusters; \S 6).
The velocity distribution of the matter, however, fits a
Maxwellian only for the low-density CDM model (where it is similar
to the velocity distribution of the clusters;  Figures 5b, 6b).
The matter velocity distribution in the $\Omega=1$ models can
not be represented by a Maxwellian - it exhibits a more extended
tail at high velocities (Figures 5a, 6a). This tail is contributed
mostly by the high velocity-dispersion matter within rich
clusters in this model (\S 4); it is best approximated by an
exponential fit (Cen \& Ostriker 1993).

The HDM and PBI models yield similar results to those of the
$\Omega=1$ CDM:  the clusters are well fit by Maxwellian distributions,
but the matter velocities exhibit more extended tails than expected
in a Maxwellian distribution.

A Maxwellian distribution of the large-scale velocity field is expected
from Gaussian initial density fluctuations. The above results suggest
that clusters of galaxies, with their Maxwellian velocity distribution,
provide a fundamental tracer of the large-scale velocity field. The
cluster velocities can test the type of the initial density field
(Gaussian or non-Gaussian).

\bsn
\centerline{6. PECULIAR VELOCITIES OF CLUSTERS IN SUPERCLUSTERS}
\msn

Approximately 10\% of all rich clusters of galaxies in the models
exhibit large peculiar velocities, $v \ge 10^3$ km s$^{-1}$ (\S 3).
What is the origin of these large velocities?  Do they originate
from the gravitational interaction in dense superclusters (or close
cluster pairs) as suggested by Bahcall, Soneira \& Burgett (1986)?
We investigate this question below.

We identify superclusters (groups of clusters of galaxies) in each
of the model simulations using a linkage algorithm for the rich
clusters (\S 2,3). A linkage-length of $10$h$^{-1}$ Mpc is used;
it identifies dense superclusters. All rich cluster pairs separated
by less than $10$h$^{-1}$ Mpc are grouped into a ``supercluster''.
The superclusters contain, by definition, two or more rich clusters.
This procedure follows the standard observational selection of
superclusters (e.g., Bahcall and Soneira 1984).

The velocity distribution of clusters that are supercluster members
is determined for each model. This distribution is compared with the
velocity distribution of {\it all} clusters (i.e., not just those that are
supercluster members), as well as of the {\it isolated} clusters (i.e.,
{\it not} in superclusters).

The results are presented in Figure 7. The velocity distributions
of isolated clusters, clusters in superclusters, and all clusters are
compared with each other. Figures 7a-b present the integrated
velocity distribution for the $\Omega=1$ and $\Omega=0.3$ CDM models.
The results show, as expected, that the velocity distribution of
clusters in dense superclusters differs somewhat from that of
isolated clusters. For $\Omega=1$ CDM, only $\sim 2$\% of all
isolated clusters have velocities $\ge 10^3$ km s$^{-1}$;
however, $\sim 30$\% of supercluster members exhibit these same
high velocities. The effect is similar, though somewhat smaller
in magnitude, in the low-density CDM model.

\bsn
\centerline{7.  COMPARISON WITH OBSERVATIONS}
\msn

Observational determination of peculiar velocities of galaxies, groups,
and clusters is difficult, since the true distances of the objects
and hence their Hubble velocities are uncertain. However, data are
available for the peculiar velocities of some samples of groups
and of clusters of galaxies. We use the group and cluster peculiar
velocities measured by Aaronson \etal (1986), Mould \etal
(1991, 1993), and Mathewson \etal (1992), who employ the
Tully-Fisher (TF) method for distance indicators, and
Faber \etal (1989), who use the $D_n - \sigma$ method. Groups with
observational velocity uncertainties $\ge 900$ km s$^{-1}$ are
excluded from the analysis. A total of 48 group and cluster
peculiar velocities are available from the TF method, and 91 from
$D_n - \sigma$. The total sample of peculiar velocities includes
123 non-overlapping groups and clusters. These data are used to
determine the observed velocity distribution of groups of galaxies.
The data correspond to low-threshold groups, with populations
comparable to the simulated groups studied above. As seen in
\S 3-4, the velocity distribution is insensitive to the group
threshold: small groups and rich clusters generally yield similar
results.

The observed differential and integrated group velocity distributions
are presented in Figures 8 to 10.  Here we use the one-dimensional
velocities, as observed ($v_{1D}$); they are compared with the 1-D
velocities in the models (as opposed to the 3-D velocities used in the
previous sections, since the actual 3-D velocities can not be
determined observationally). We present in different symbols the
data obtained from the TF, $D_n - \sigma$, and total (combined)
samples (Figures 8a - 10a). The results from the different
sub-samples appear to be consistent with each other; the total
sample can thus be used for comparison with model expectations.

The number of rich clusters in the observed sample is rather small
($\sim 18$ $R \ge 0$ clusters). Within the large statistical
uncertainties of such a small sample, the observed rich cluster velocity
distribution is consistent with that of the groups, and thus consistent
with the model comparisons discussed below.

The observed velocity distribution is superimposed on the 1-D velocity
distribution expected for groups for each of the four models
(dotted line; Figs. 8a-d to 10a-d). The shape of the functions
differs from those of Figures 3-5 since the 1-D velocity distribution
is plotted instead of the 3-D. The 3-D distribution is proportional
to $v^2$ at small velocities; instead, the 1-D velocities
exhibit a Gaussian distribution, as expected for a 3-D Maxwellian.
In comparing model expectations with observations, the model
velocity distribution (dotted line) is convolved with the observational
uncertainties; each model cluster is given an uncertainty drawn at
random from the actual distribution of observed uncertainties. The
convolved model distribution of groups is shown by the dashed lines in
Figs. 8-10; the convolved rich cluster distributions are shown by
the solid lines, for comparison. The convolution flattens the model
distributions, as expected, and produces a high-velocity tail. The
convolution also reduces the differences between the different model
distributions, as well as between groups and clusters.

A comparison between the data and the convolved models suggests that
the observed and model velocity distributions are generally consistent
with each other. The observations exhibit a long tail of high velocity
groups and clusters, to $v_{1D} \sim 3000$ km s$^{-1}$ (seen most
clearly in Figures 9-10). This
high-velocity tail is most consistent with the HDM model predictions
(Fig. 9c, 10c). However, due to the large observational uncertainties,
the data are also consistent with the other models
(at a $\le 2\sigma$ level). A K-S test of the velocity
distribution indicates that all models are consistent with the data
at a significance level of $\sim 10$\%-20\% (with PBI showing a lower
significance level of 3\%). The model fits with observations may be
improved if a large fraction of the observed groups and clusters are
located in dense superclusters (\S 6; Fig. 7).
Since the current observational uncertainties are large, and the
effect of model convolution is strong (in fact it yields most of
the high-velocity tail), more accurate velocity data are needed in
order to further constrain the cosmological models. One effect of
large observational errors is to produce an artificial high-velocity
tail. It seems likely that more accurate cluster velocities will be
smaller than currently suggested by the observed high-velocity tail
of Figures 9-10.

The HDM model provides the best fit to the observed high-velocity tail;
however, this predicted tail is caused by the high normalization of
the model ($\sigma_8=1$; Table 1), which is $2\sigma$ above
the COBE normalization. (A high normalization is needed in order to
minimize the problem of late galaxy formation.) A lower normalization,
as required by COBE, will reduce the high velocity tail, making it
comparable to the $\Omega=1$ CDM. The $\Omega=1$ CDM model has
similar problems; its high COBE normalization used here ($\sigma_8=1$)
is inconsistent with the mass-function and correlation-function of
clusters of galaxies, and with the small-scale pair-wise galaxy
velocities (Bahcall \& Cen 1992, Kofmann \etal 1993, Ostriker 1993).
A lower normalization will yield a lower velocity tail than
given above (and will still be inconsistent with the cluster
correlations and mass-function). An $\Omega=1$ mixed HDM+CDM model
(30\% + 70\%, respectively), properly normalized to COBE
($\sigma_8 \sim 0.67$), will yield a velocity distribution comparable
to that of the $\Omega \sim 1$ CDM models. Finally,
the COBE-normalized low-density CDM model, which is consistent with
most large-scale structure observations (\S 4), is also consistent
(at $\sim 20$\% significance level, K-S test) with the observed
peculiar velocity distribution of  groups and clusters studied above.

\bsn
\centerline {8. CONCLUSIONS}
\msn

The distributions of peculiar velocities of rich clusters of galaxies,
of groups, and of the total matter have been investigated using
large-scale simulations of four cosmological models:  standard
CDM ($\Omega=1$), low-density CDM, HDM ($\Omega=1$), and PBI.
The main conclusions are summarized below.

\item {(1)} Rich clusters of galaxies exhibit a robust, Maxwellian
distribution of peculiar velocities in all models studied. The
distribution peaks at $v \sim 500$ km s$^{-1}$, and extends to
high velocities of $v \sim 1500$ km s$^{-1}$. The velocity
distribution is similar in all the models (with the PBI distribution
shifted to somewhat lower velocities). Approximately 10\% of all
rich clusters move with peculiar velocities of
$v \ge 10^3$ km s$^{-1}$ (somewhat lower for PBI).
\item {(2)} The highest velocity clusters, with  $v > 10^3$ km s$^{-1}$,
originate frequently in dense superclusters.
\item {(3)} The velocity distribution of model clusters is insensitive
to the cluster selection threshold (i.e., richness). The velocity
distribution of small groups of galaxies is similar to that of the
rich clusters, with only a minor suggested shift to larger velocities
(larger shift for HDM).
\item {(4)} The velocity distribution of the total matter is similar
to the velocity distribution of groups and of rich clusters for the
low-density CDM model. In all other models, especially those with
$\Omega=1$, the total mass exhibits a larger tail of high-velocities
($\ge 2000-3000$ km s$^{-1}$ ) than do the clusters. This high-velocity
tail of the mass distribution reflects the large velocity-dispersions
that exist within rich clusters of galaxies in high-density models.
The mass velocity distribution for the $\Omega=1$ models can not be
described by a Maxwellian; it is better approximated by an exponential.
The low-density CDM model, on the other hand, exhibits velocity
distributions of clusters, of groups and of the total matter that
are all similar to each other, and that are each described well by a
Maxwellian distribution.
\item{(5)} The observed distributions of peculiar velocities of groups
and of clusters of galaxies are generally consistent with the model
predictions when convolved with the observational uncertainties. The
observed velocities exhibit a large tail of high-velocity groups and
clusters, to $v_{1D} \sim 3000$ km s$^{-1}$. This high-velocity tail
is most consistent with the HDM model; however, the overall observed
velocity distribution is consistent with all the models.
\item{(6)} There are large uncertainties in the existing measurements
of peculiar velocities. More accurate peculiar velocity observations
of groups and of clusters of galaxies are likely to yield a lower
velocity-tail than suggested by existing observations.
\item{(7)} Future observations of peculiar velocities of clusters
and groups of galaxies can help to constrain cosmological models
and to test whether the initial density field was Gaussian or
non-Gaussian.

In summary, we conclude that groups and clusters of galaxies provide
robust and efficient tracers of the large-scale peculiar velocity
distribution. The current data are consistent with the cluster
velocity distribution expected from the four models studied. In
particular, the observed velocity distribution is consistent
with a COBE-normalized low-density CDM-type model, which
best fits other large-scale structure observations. It is also
consistent with an initial density field that is Gaussian.

\bsn
\centerline {ACKNOWLEDGMENTS}
\msn

It is a pleasure to acknowledge NCSA for allowing
us to use their Convex-3880 supercomputer,
on which our computations were performed.
We gratefully thank J. Goodman, J.R. Gott, J.P. Ostriker,
D.N. Spergel and R. Wijers for useful discussions.
R.Y.C is supported by
NASA grant NAGW-2448 and NSF grants AST91-08103.
M.G. is supported in part by NASA grant
NAGW-2448 and NSF grant AST90-20506.
\vfill\eject
\centerline {REFERENCES}
\msn
\smallskip
\refset
Aaronson, M., Bothun, G., Mould, J., Huchra, J., Schommer, R.A., \&
Cornell, M.E. 1986, ApJ, 302, 536
\smallskip
\refset
Bahcall, N.A., \& Cen, R. 1992, ApJ, 398, L81
\smallskip
\refset
Bahcall, N.A., \& Cen, R. 1993, ApJ, 407, L49
\smallskip
\refset
Bahcall, N.A. \& Soneira, R.M. 1983, ApJ, 270, 20
\smallskip
\refset
Bahcall, N.A. \& Soneira, R.M. 1984, ApJ, 277, 27
\smallskip
\refset
Bahcall, N.A., Soneira, R.M., \& Burgett, W. 1986, ApJ, 311, 15
\smallskip
\refset
Bahcall, N.A., \& West, M. 1992, ApJ, 392, 419
\smallskip
\refset
Bardeen, J.M., Bond, J.R., Kaiser, M., \& Szalay, A.S. 1986, ApJ, 304, 15
\smallskip
\refset
Bertschinger, E., \& Dekel, A. 1989, ApJ, 336, L5
\smallskip
\refset
Burstein, D., Davies, R.L., Dressler, A., Faber, S.M., Lynden-Bell, D.,
Terlevich, R., \& Wegner, G. 1987, in "Galaxy Distances and Deviations
from Universal Expansion," ed. B. Madore \& R.B. Tulley
(Dordrech: Reidel), 123
\smallskip
\refset
Cen, R.Y. 1992,  ApJS, 78, 341
\smallskip
\refset
Cen, R., \& Ostriker, J.P. 1992, ApJ, 399, 331
\smallskip
\refset
Cen, R., \& Ostriker, J.P. 1993, ApJ, 417, 415
\smallskip
\refset
Cen, R., Ostriker, J.P., \& Peebles, P.J.E. 1993, ApJ, 415, 423
\smallskip
\refset
Cen, R., Gnedin, N.Y., \& Ostriker, J.P. 1993, ApJ, 417, 387
\smallskip
\refset
Croft, R.A.C., \& Efstathiou, G. 1993, preprint
\smallskip
\refset
Dalton, G.B., Efstathiou, G., Maddox, S., \& Sutherland, W. 1992,
ApJ, 390, L1
\smallskip
\refset
Dekel, A. 1994, ARAA, 32
\smallskip
\refset
Dressler, A., Faber, S.M., Burstein, D., Davies, R., Lynden-Bell, D.,
Terlevich, R., \& Wegner, G. 1987, ApJ, 313, L37
\smallskip
\refset
Einasto, M., Tago, E., \& Einasto, J. 1994, MNRAS, submitted
\smallskip
\refset
Faber, S.M., Wegner, G., Burstein, D., Davies, R.L., Dressler, A.,
Lynden-Bell, D., \&  Terlevich, R.J. 1989, ApJS, 69, 763
\smallskip
\refset
Klypin, A.A., \& Kopylov, A.I. 1983, Soviet Astron. Lett., 9, 41
\smallskip
\refset
Kofman, L., Bertschinger, E., Gelb, J.M., Nusser, A., \& Dekel, A.
1993, ApJ, submitted
\smallskip
\refset
Kofman, L., Gnedin, N., \& Bahcall, N.A, 1993, ApJ, 413, 1
\smallskip
\refset
Lauer, T., \&  Postman, M. 1993, preprint
\smallskip
\refset
Maddox, S.J., Efstathiou, G., Sutherland, W., \& Loveday, J. 1990,
MNRAS, 242, 43
\smallskip
\refset
Mathewson, D.S,, Ford, V.L., \& Buchhorn, M. 1992, ApJS, 81, 413
\smallskip
\refset
Mould, J.R., Staveley-Smith, L., Schommer, R.A., Bothun, G.D.,
Hall, P.J., Han, M., Huchra, J.P., Roth, J.,
Walsh, W., \& Wright, A.E. 1991, ApJ, 383, 467
\smallskip
\refset
Mould, J.R., Akeson, R.L., Bothun, G.D., Han, M., Huchra, J.P.,
Roth, J., \& Schommer, R.A. 1993, ApJ, 409, 14
\smallskip
\refset
Nichol, R.C., Collins, C.A., Guzzo, L., \& Lumsden, S.L. 1992,
MNRAS, 255, 21
\smallskip
\refset
Ostriker, J.,P. 1993, ARAA, 31, 689
\smallskip
\refset
Peacock, J.A., \& West, M. 1992, MNRAS, 259, 494
\smallskip
\refset
Postman, M., Huchra, J., \& Geller, M. 1992, ApJ, 384, 404
\smallskip
\refset
Ramella, M., Geller, M., \& Huchra, J. 1989, ApJ, 344, 57
\smallskip
\refset
Rhoads, J., Gott J.R., \& Postman, M., 1994, ApJ, 421, 1
\smallskip
\refset
Rubin, V.C., Thonnard, N., Ford, W.K., \& Roberts, M.S. 1976, AJ, 81, 719
\smallskip
\refset
Smoot, G.F., \etal 1992, ApJ, 396, L1
\smallskip
\refset
Suto, Y., Cen, R., \& Ostriker, J.P. 1992, ApJ, 395, 1
\vfill\eject
\noi
\centerline {FIGURE CAPTIONS}
\bsn

FIGURE 1. Integrated 3-D peculiar velocity distribution of rich
clusters of galaxies in four cosmological models: $\Omega=1$ CDM
(solid line), $\Omega=0.3$ CDM (dashed line), $\Omega=1$
HDM (dash-dot line), and $\Omega=0.3$ PBI (dotted line).
\bsn
FIGURE 2. Differential 3-D peculiar velocity distribution of rich
clusters of galaxies for the four cosmological models of Fig. 1
(same notation).
\bsn
FIGURE 3. Differential velocity distribution (3-D) of rich clusters
(solid line), of groups (dashed line), and of the total matter (dotted
line), for different models:  (a) $\Omega=1$ CDM; (b) $\Omega=0.3$ CDM;
(c) $\Omega=1$ HDM; and (d) $\Omega=0.3$ PBI.
\bsn
FIGURE 4. Same as Figure 3 but for the integrated velocity distribution
(of rich clusters, groups, and matter).
\bsn
FIGURE 5. Differential velocity distribution (3-D) of clusters
and of the total matter in the models
(dark solid and dotted lines, respectively), and their
comparison with a Maxwellian distribution (faint lines) (\S 5).
(a) $\Omega=1$ CDM; (b) $\Omega=0.3$ CDM.
\bsn
FIGURE 6. Same as Figure 5 but for the integrated velocity
distribution.
\bsn
FIGURE 7. Integrated velocity distribution (3-D) of model clusters
that are:  members of dense superclusters (dashed line); isolated
clusters (dotted line); and all clusters (solid line) (\S 6).
(a) $\Omega=1$ CDM; (b) $\Omega=0.3$ CDM.
\bsn
FIGURE 8. Comparison of observations and model simulations.
The observed differential velocity distribution (in 1-D velocities)
of groups of galaxies as determined from Tully-Fisher (TF) and
$D_n - \sigma$ distance-indicators (stars and open circles,
respectively), and for the combined sample (dark circles with
$\sqrt N$ uncertainties indicated) are presented (\S 7).
The model 1-D group velocity distribution (dotted line), and its
convolved distribution (dashed line; convolved with the observational
velocity uncertainties) are shown. The convolved distribution
of simulated rich clusters (faint solid line) is also shown,
for comparison. The observations should be compared with the
convolved model group simulations (dashed line).
(a) $\Omega=1$ CDM, (b) $\Omega=0.3$ CDM;
(c) $\Omega=1$ HDM; (d) $\Omega=0.3$ PBI.
\bsn
FIGURE 9. Same as Figure 8, but in a logarithmic scale.
\bsn
FIGURE 10. Same as Figure 9, but for the integrated velocity
distributions. (The $\sqrt N/N$ of each bin is presented, for
illustration only, by the vertical bars).
\vfill\eject
\hsize=4.5truein
\hoffset=2.20truecm
\vglue 1cm
\centerline {TABLE 1}
\vskip 2mm
\centerline {MODEL PARAMETERS}
\bsn
$$\vbox{\tabskip=0.35truecm
\halign to \hsize {# \hfil &\hfil# \hfil &\hfil#\hfil &\hfil#\hfil
&\hfil# \hfil &\hfil# \hfil&\hfil# \hfil &\hfil#\hfil &\hfil#\hfil
&\hfil# \hfil \cr
\noalign {\bigskip}
\noalign{\hrule height 0.4pt}
\noalign{\smallskip}
\noalign{\hrule height 0.4pt}
\noalign{\bigskip}
MODEL& & &$\Omega$& &$\Omega_\Lambda$& &$h$& &$\sigma_8$ \cr
\noalign {\bigskip}
\noalign{\hrule height 0.4pt}
\noalign{\bigskip}
CDM & & & 1.0 & & 0.0 & & 0.50 & & 1.05 \cr
\noalign {\bigskip}
CDM & & & 0.3 & & 0.7 & & 0.67 & & 0.67 \cr
\noalign {\bigskip}
HDM & & & 1.0 & & 0.0 & & 0.75 & & 1.00 \cr
\noalign {\bigskip}
PBI & & & 0.3 & & 0.0 & & 0.80 & & 0.80 \cr
\noalign {\bigskip}
\noalign{\hrule height 0.4pt}
\noalign {\bigskip}
}}$$
\par\vfill\eject

\bye